\documentclass{article}
\usepackage[papersize={210mm,297mm}, total={160mm,250mm}]{geometry}

\usepackage[square,numbers]{natbib}
\usepackage[utf8]{inputenc} 
\usepackage[T1]{fontenc} 
\usepackage{hyperref} 
\usepackage{url} 
\usepackage{booktabs} 
\usepackage{amsfonts} 
\usepackage{nicefrac} 
\usepackage{microtype} 
\usepackage{lipsum}
\usepackage{fancyhdr} 
\usepackage{graphicx} 
\graphicspath{{media/}} 
\usepackage{xcolor}
\usepackage[para]{threeparttable}

\usepackage{amsmath}
\usepackage{nccmath}
\usepackage{tabularx}
\usepackage{multirow}
\usepackage{amssymb}
\usepackage{subfig}

\usepackage{soul}
\usepackage{color}
\sethlcolor{white}
\usepackage{listings}

\graphicspath{ {images/} }
\title{ Chat2VIS: Fine-Tuning Data Visualisations using Multilingual Natural Language Text and Pre-Trained Large Language Models

}

\author{
Paula Maddigan\footnote{\href{mailto:t.susnjak@massey.ac.nz}{P. Maddigan Email: Paula.Maddigan.1@uni.massey.ac.nz}}  
and 
  Teo Susnjak\footnote{\href{mailto:t.susnjak@massey.ac.nz}{T. Susnjak Email: t.susnjak@massey.ac.nz}} 
\\
  School of Mathematical and Computational Sciences \\
  Massey University \\
  Auckland, New Zealand \\
}

\begin{document}
\maketitle

\begin{abstract}
The explosion of data in recent years is driving individuals to leverage technology to generate insights.  Traditional tools bring heavy learning overheads and the requirement for understanding complex charting techniques. Such barriers can hinder those who may benefit from harnessing data for informed decision making.  The emerging field of generating data visualisations from natural language text (NL2VIS) addresses this issue. This study showcases Chat2VIS, a state-of-the-art NL2VIS solution. It capitalises on the latest in AI technology with the upsurge in pre-trained large language models (LLMs) such as GPT-3, Codex, and ChatGPT. Furthermore, the rise in natural language interfaces (NLI) and chatbots is taking centre stage. This work illustrates how Chat2VIS leverages similar techniques to fine-tune data visualisation components beyond that demonstrated in previous approaches. In addition, 
this paper presents the flexibility of Chat2VIS to comprehend multilingual natural language requests. No other NL2VIS system has demonstrated this unique talent. In concluding, this research provides quantitative benchmarking evaluations to contribute to the paucity of NL2VIS standards.

\end{abstract}



\maketitle

\section{Introduction}\label{introduction}
The volume of data accumulated within organisations is experiencing rapid growth. Industries are becoming increasingly reliant on data insights to aid in their decision making. Data visualisations offer an effective and compelling approach for communicating valuable information and facilitating informed business decisions.  The aim of improving usability  across a broader range of users\cite{luo2021natural} to leverage visualisation tools for data insights has led to the emerging field of Natural Language Interfaces (NLI). 

The notion of expressing visualisation requests using natural language (NL) text and a user-friendly interface is becoming a sought after goal\cite{shen2021towards}. To generate suitable charts without the understanding of chart type nuances is an enticing objective.
Well designed NLIs can avoid the need for programming skills and arduous learning curves\cite{wang2022towards}. 

Recently, the surge in interest of large language models (LLMs) is driving research to develop NLIs leveraging this technology. Using OpenAI's GPT-3, Codex, and ChatGPT models, this study explores the degree LLMs can be  capitalised on.  
These models are trained on a large amount of internet information and code repositories.  Consequently, they exhibit a high level of skill in language semantics and code scripting. This unique capability is of significant importance in this work. 

One remarkable feature of the conversational capabilities of ChatGPT is its unique ability to build on prior exchanges.  Incorporating this feature within NLIs can facilitate the fine-tuning of visualisations. This study demonstrates how iterative enhancements to user prompts can further define plot elements, beyond those of previously developed NL2VIS architectures. 

Primarily the LLMs were trained using the English language and a selection of major European languages.  In addition a large corpus of text in other languages was included. The proficiency of the LLMs in understanding and responding to requests in such languages is dependent on the quantity and quality of training data available. There is little evidence of natural language to visualisation (NL2VIS) studies exploring the capabilities of generating visualisations from queries in languages other than English. One such explanation is  the shortage of multilingual natural language datasets available to train existing NL2VIS models. The NLI explored in this study demonstrates how the innovative architecture capitalises on the LLMs multilingual skills to interpret requests in mixed languages and render the intended visualisations.

It is crucial to establish benchmarks to facilitate advancements in this area. Methodically and systematically assessing the efficacy of new approaches provides a robust measure of improvement in NL2VIS systems.
Presented in this work is a unique framework fabricated in an effort to address this growing concern.  Nevertheless, the degree of accuracy is to some extent contingent on the effectiveness of the proposed methodology for comparing visualisations, rather than solely a reflection of the NL2VIS system performance.  Consequently it should be pertinent to emphasise the process to gauge accuracy is not definitive, but rather measurable on a scale of correctness. To conclude we highlight the importance of ensuring benchmarks can fulfil all the quality characteristics \cite{Kounev2020Benchmark} 
of \textit{reproducibility}, \textit{fairness}, and \textit{verifiability} expected of such baselines. We note it is imperative the benchmark embodies a measurement methodology, defining the process to implement the standard, collect measurements, and evaluate the results \cite{Kounev2020Benchmark}.

\subsection{Contribution}
This study advances the field of NL2VIS by presenting the novel features within the LLM-based NLI Chat2VIS.
We demonstrate its extraordinary capacity to comprehend multilingual natural language texts to generate data visualisations, establishing it as a truly global tool. We illustrate its remarkable power to incorporate iterative refinements to queries to polish the generated charts and aesthetics. We show with its novel design it is not confined to solely refining a predefined set of chart components as demonstrated by previous NL2VIS approaches. 

The current state of benchmarking tools for accurately measuring the capabilities of NL2VIS is still in its infancy. We contribute to existing literature 
by presenting an innovative automated approach to conducting quantitative analysis of NL2VIS performance. Specifically, we present the results of our evaluation of Chat2VIS against two benchmarks.  Our evaluation  highlights the challenges of developing robust structured methodologies and measurable baseline standards for NL2VIS.

\section{Related Work}
Early NL2VIS systems were built on symbolic-based NLP approaches, relying on heuristic algorithms \cite{liu2020extracting}, rule-based architectures, and probabilistic grammer-based methods for translating NL queries.  Although each technique displayed increasing accuracy, they required more computational resources.  Systems such as Articulate \cite{sun2010articulate}, DataTone \cite{gao2015datatone}, Eviza \cite{setlur2016eviza}, and Deep-Eye \cite{qin2018deepeye} all used varying symbolic NLP methodologies in translating NL to data visualisations. However notable approaches like NL4DV \cite{narechania2020nl4dv} and FlowSense \cite{yu2019flowsense} employed NLTK \cite{loper2002nltk}, NER, and Stanford CoreNLP \cite{manning2014stanford} semantic parsers to improve accuracy.  

\begin{figure}[!h]
\centering
\includegraphics[width=3in]
{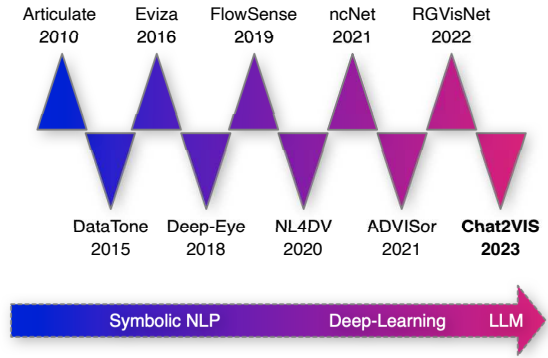}
\caption{NL2VIS timeline illustrating the evolution of NL2VIS systems.}
\label{nl2vis_timeline}
\end{figure}

Recent advancements in NL2VIS have focused on deep-learning models to achieve greater levels of adaptability, robustness and flexibility compared to that of previous approaches\cite{voigt2021challenges}. 
Systems such as  ADVISor \cite{liu2021advisor} are supported by BERT \cite{devlin2018bert}, a large language transformer-based model. The rendered visualisation styles are predetermined based on a defined mapping rule.  

 An alternative transformer-based approach, ncNet \cite{luo2021natural}, is a machine learning model  trained using the nvBench \cite{luo2021nvbench} dataset.  The model accepts an optional chart template in addition to the requested NL query to guide chart styling of the rendered visualisation. The system has recently been expanded to include speech-to-visualisation capabilities \cite{tang2022sevi}.
 
 Futhermore, the hybrid approach of RGVisNet \cite{song2022rgvisnet} initially retrieves the most relevant visualisation query from a large-scale visualisation codebase.  It then revises it via a GNN-based deep-learning model, and subsequently generates the visualisation.  

 The evolution of NL2VIS systems are illustrated in Fig. \ref{nl2vis_timeline}, depicting the transition from symbolic NLP into deep learning approaches. The latest state-of-the-art artifact, Chat2VIS \cite{maddsus2023chat2vis}, presents the first NL2VIS NLI to generate data visualisations via LLMs. It addresses the next generation of NL2VIS architecture, simplifying the NL2VIS pipeline. The underlying structure provides flexibility and robustness around free-form and complex visualisation requests. Decisions pertaining to suitable chart selection and aesthetics are delegated to the LLMs.  
 
 The architecture underpinning Chat2VIS is exceptionally flexible and decidedly diverse enough to further refine charting elements using NL without additional enhancements to the NL2VIS architecture. This is the first study to address this gap evident in earlier systems. In addition, unseen in previous approaches, this work demonstrates the art of fulfilling multilingual requests with ease, omitting the need for additional prompting, further architectural manipulations, or model retraining.  

With the sparse existence of NL2VIS benchmarks, we seek to evaluate Chat2VIS against the only two baselines nvBench \cite{luo2021nvbench} and the NLV utterance corpus \cite{Srinivasan2021Utt} identified in existing literature. Evaluations \cite{song2022rgvisnet}\cite{Srinivasan2021Utt} against these benchmarks for current NL2VIS approaches provide a degree of comparison for this study. To that end our analysis contributes to the gap distinctly evident in the literature regarding NL2VIS benchmarking.

\section{NL2VIS Architecture}

Chat2VIS generates data visualisations using NL free-form text.  
With an interface utilising OpenAI's state-of-the-art LLMs, it demonstrates unique decision making skills to autonomously select chart types and plot elements.  

\subsection{Large Language Models}
Chat2VIS is based on the Davinci family of models, currently the most capable and advanced model set available within the OpenAI suite. It employs GPT-3 model \textit{"text-davinci-003"}, Codex model \textit{"code-davinci-002"}, and contrasts results with the latest state-of-the-art ChatGPT model \textit{"gpt-3.5-turbo"}.

Using OpenAIs text completion endpoint API to access Codex and GPT-3 models, 
it retains default parameters with the exception of adjustments to the following: 
\begin{enumerate}
    \item Setting of the temperature parameter to zero to encourage the LLMs to be more consistent with their code generation — causing them to use more-common syntax with less creativity;
    \item Evading excessively verbose scripts by setting the  max\_tokens parameter to 500 — an ample limit for this study; and
    \item Requesting a stop parameter of \textit{"plt.show()"}. This will cease generation upon plot rendering syntax - avoiding the LLMs presenting alternative scripts. 
\end{enumerate}

\begin{figure}[!h]
\centering
\includegraphics[width=3in]
{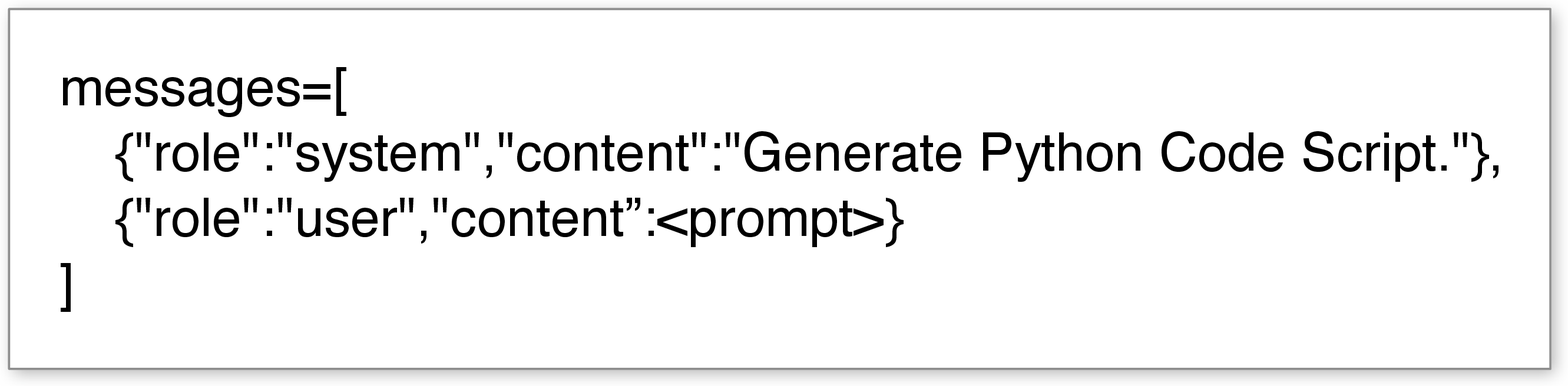}
\caption{ChatGPT API message structure.}
\label{chatgpt_api}
\end{figure}

With the recent release of the official ChatGPT API, 
it uses the new chat-completion endpoint\footnote{https://openai.com/blog/introducing-chatgpt-and-whisper-apis}. Traditionally, prompts submitted to Codex and GPT-3 models are depicted as a succession of "tokens".  In contrast, requests to ChatGPT are submitted as a sequence of messages, and subsequently converted to tokens using the new "Chat Markup Language"  (ChatML)\footnote{https://github.com/openai/openai-python/blob/main/chatml.md}. Fig. \ref{chatgpt_api} shows the message structure.

\subsection{Chat2VIS}

\begin{figure*}[!h]
\centering
\includegraphics[width=\textwidth]
{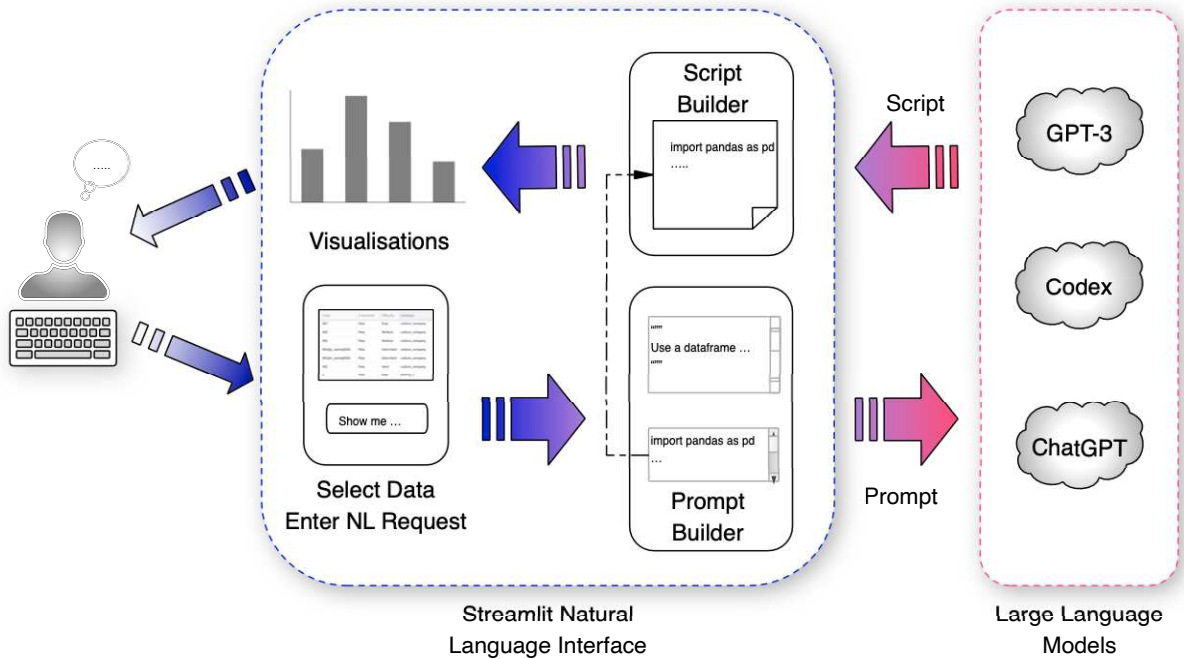}
\caption{The Chat2VIS architecture  translating natural language text into data visualisations via large language models.}
\label{chat2vis_overview}
\end{figure*}
Chat2VIS\footnote{https://chat2vis.streamlit.app/} is fabricated using Streamlit, an open-source Python framework for building web-based applications. The adoption of  this technology provides a means to encapsulate the NL2VIS user interface, prompt engineering, LLM connectivity, and rendering of visualisations from generated scripts.
Fig. \ref{chat2vis_overview} depicts an overview of the architecture of Chat2VIS. Using the interface illustrated in  Fig. \ref{interface}, a user enters a request in the form of natural language text in reference to a selected dataset.  Chat2VIS engineers the prompt, submits it to the chosen LLMs, formats the returned script, and renders the visualisation.  

\begin{figure}[!h]
\centering
\includegraphics[width=3in]
{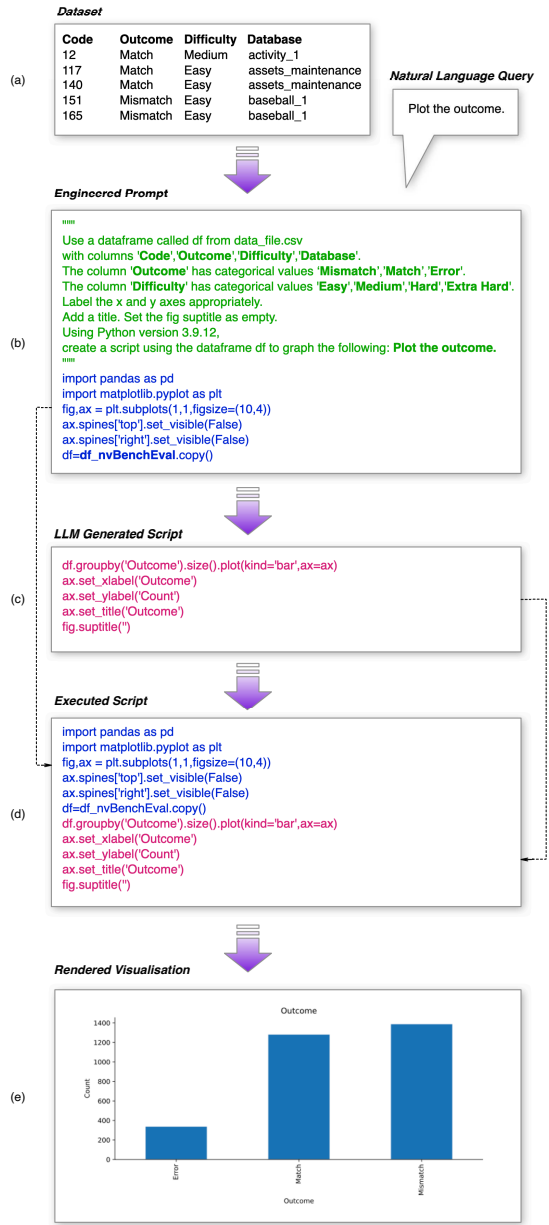}
\caption{Illustrated example of the process to convert a NL query into a data visualisation.}
\label{chat2vis_eg}
\end{figure}

Fig. \ref{chat2vis_eg} illustrates the inner workings of Chat2VIS using Codex.  The architecture is discussed by way of an example dataset 
created from the results of our benchmarking evaluation in Section  \ref{section:nvBench_eval}.  The process is described as follows:
 
\begin{enumerate}
    \item Fig. \ref{chat2vis_eg}(a) shows a sample of the dataset 
    together with the query \textit{"Plot the outcome."}.
    \item The engineered prompt in Fig. \ref{chat2vis_eg}(b), described in further detail in Fig. \ref{chat2vis_prompt}, comprises of two parts: (1) a Python docstring description, encapsulated with triple double quotes (green); (2) a Python code section providing a starting point for the requested script (blue).  The \textbf{bolded} type highlights variable substitution values specific to this example.
    \item Upon submission of the prompt to the selected LLMs, Fig. \ref{chat2vis_eg}(c) details the returned script (red) — a continuation of the code section within the engineered prompt.
    \item 
    In Fig. \ref{chat2vis_eg}(d), the prompt code section (blue) is inserted at the beginning of the generated script (red).
    \item The newly-created script is executed to render the requested visualisation, as pictured in Fig. \ref{chat2vis_eg}(e). 
\end{enumerate}

\section{Methodology}
\label{sec:methodology}
This study explores the refinement of plot aesthetics using natural language queries, and the flexibility of Chat2VIS in comprehending multilingual requests. 

Furthermore, we conduct a quantitative evaluation against two benchmark datasets developed in previous studies.

\subsection{Chart Refinements and Multilingual Requests  }
Previous work \cite{maddsus2023chat2vis} confirms the unique decision-making skills of the LLMs to autonomously select suitable chart types and plot aesthetics.
Using Chat2VIS, we demonstrate 
the efficacy of iterative refinements to the input query for nominating a specific chart type, enhancing plot elements, and changing styles. 

We further illustrate the proficiency of the LLMs to interpret multilingual requests to generate data visualisations.   We explore the flexibility in chart labeling directed by language preference. 

We present four case studies showcasing these refinements using Chat2VIS and the underlying LLMs to render requested visualisations via natural language. 
The results are assessed visually for accuracy and compliance.

\subsection{Quantitative Evaluation}
We conduct a more comprehensive quantitative analysis using the nvBench benchmark\footnote{https://sites.google.com/view/nvbench}.
Encompassing 153 databases, 7,274 visualisations, and 7 chart types, it is considered the first public large-scale NL2VIS benchmark \cite{luo2021nvbench}.   Each example instance comprises of a natural language-to-visualisation pair, denoted (NL, VIS).  Attributes are stored inside a json specification, permitting Vega-Lite chart rendering. Examples are further classified into four categories to denote the difficulty of the query — easy, medium, hard, and extra hard.

In addition, we perform a second evaluation using the NLV Utterance  dataset\footnote{https://nlvcorpus.github.io/} 
\cite{Srinivasan2021Utt}, referred to in our study as nlvUtterance. The benchmark covered three databases\footnote{https://github.com/TsinghuaDatabaseGroup/nvBench/databases.zip}: movies, cars, and superstore. 
 This benchmark comprises of 814 natural language queries, with 10 visualisations for each database. Queries were generated from the results of an online study using 102 participants suggesting utterances for the display of each respective chart. 

\subsubsection{Model Selection}
We select the Codex \textit{"code-davinci-002"} model to measure results against nvBench. Codex, evolved from GPT-3, was trained on an immense amount of publicly-available GitHub code. It is skilled in more than a dozen programming languages, most notably Python, the underlying programming language of Chat2VIS.  Codex is available in Davinci or Cushman models.  Among the OpenAI suite of models, the Davinci family is the most capable, and can often perform all tasks of other models using fewer instructions.
Cushman, although faster, is less competent than Davinci. In prioritising accuracy over speed, the Davinci model was regarded as the most appropriate choice for this task. Therefore, we deemed Codex \textit{"code-davinci-002"} well-suited for this evaluation. 

\subsubsection{nvBench Benchmark Evaluation}
Determining how to assess the equality between a Chat2VIS chart and its nvBench counterpart is challenging. The benchmark specification omits guidance of any measurement methodology. With the plethora of dissimilarities between both visualisations, our attempts to employ image comparison tools proved unproductive.  Highlighting of differences such as axis labels, font size, and plot colouring lacked relevance, with our primary focus on the accuracy of the plotted data. Hence we settled on constructing vectors of the x and y coordinates for each plot, using these as a basis for our comparative analysis.

We select a subset of the nvBench benchmark examples for our evaluation.
 Chat2VIS is designed to generate charts from a tabular dataset. Therefore we remove nvBench instances querying multiple database tables.  This methodology is similarly adopted in the evaluation of ncNet \cite{luo2021natural}.
To exclude such examples, we can identify the SQL \texttt{JOIN} operator inside the VQL mark within the nvBench json specification.
In addition, we eliminate examples containing subqueries within the \texttt{WHERE} clause referencing tables distinct from the principal \texttt{SELECT} clause.  

\begin{figure}[!h]
\centering
\includegraphics[width=3in]
{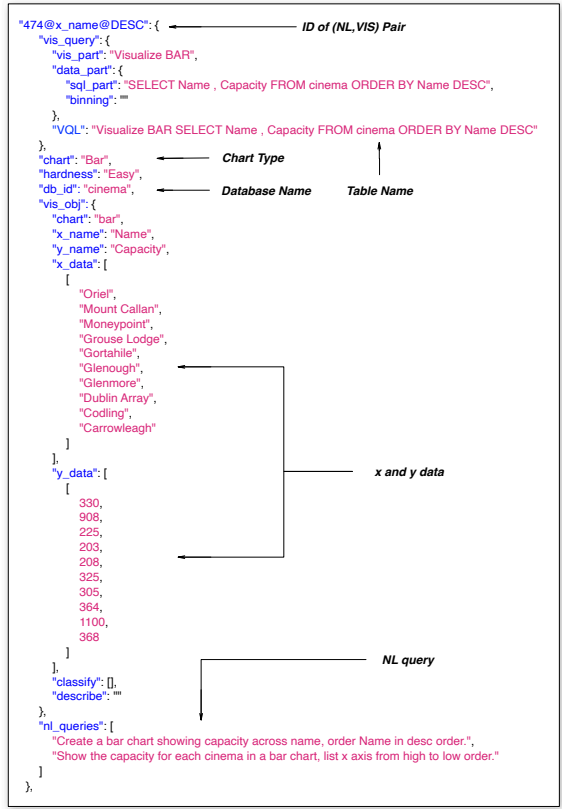}
\caption{Example json specification from nvBench.}
\label{json_eg}
\end{figure}

We further filter our test set to only include examples rendering bar charts, owing to the difficulties in establishing measurable and robust metrics to compare other chart types.  Each nvBench example generally includes more than one variation on the requested query, with all variations rendering identical visualisations. Therefore we chose to only consider the first query in each instance. Fig. \ref{json_eg} illustrates an example JSON specification\footnote{https://github.com/TsinghuaDatabaseGroup/nvBench/blob/main/NVBench.json} for the (VIS, NL) pair \textit{"474@x\_name@DESC"}, highlighting areas of interest within the specification discussed in this evaluation approach.  Covering 138 databases, the total number of nvBench examples in our test set is 3,003, with 812 considered \textit{easy}, 1572  \textit{medium}, 386 \textit{hard}, and 233 \textit{extra hard}.

Fig. \ref{testing_overview} of Appendix \ref{nvBench_method} 
summarises our testing methodology. Using an iterative process, we traverse the test set, generate Chat2VIS visualisations, compare the charts with nvBench, and categorise the results as "Match", "Mismatch", or "Error".  

\subsubsection{nlvUtterance Benchmark Evaluation}
The relatively smaller size of this test set compared to nvBench enables  us to utilise a visual non-automated approach to comparing results to Chat2VIS. 
We consider all types of charts within nlvUtterance, namely, bar charts, histograms, line charts, and scatter plots, together with their variations.  As outlined in the benchmark description\cite{Srinivasan2021Utt}: histograms and single attribute bar charts are used to visualise one categorical or quantitative attribute; bar charts, scatter plots, and line charts for two attributes; and grouped bar charts, stacked bar charts, multi-line charts, coloured scatter, and faceted scatter charts for visualising three or more attributes. 

Of note is 755 of the 814 queries are considered \textit{"singleton"} utterance sets, consisting of a single query request. The remaining 59 instances are considered \textit{"sequential"} utterance sets, containing multiple utterances with a separator character denoting a break in the queries. In examining the benchmark charts, we find two line plots void of data. Therefore we elected to omit the 56 queries associated with them, yielding 758 queries for testing.  

Chat2VIS renders charts for up to 3 models. 
Hence to capitalise on the diversity of the LLMs, we employ a three-stage methodology. Firstly, the queries are submitted to Codex and the corresponding performance metrics are presented.  Secondly, unsuccessful queries are submitted to GPT-3, with the corresponding performance again measured. Finally, any remaining mismatched queries are submitted to ChatGPT.  
The overall performance statistic for successful matches provides insight into the likelihood that a benchmark result will be generated by at least one LLM.

It is not the intention of this work to compare LLMs inter se, but instead contrast the use of LLMs with alternative approaches. Therefore, we do not present benchmark metrics comparing the performance accuracy of Codex, GPT-3 and ChatGPT relative to each other.

\section{Results}
\label{sec:results}
We demonstrate the flexibility to append to the query request to further refine the charting elements in the first three case studies. The fourth case study illustrates multilingual requests. Finally we present benchmark evaluations of Chat2VIS.

\subsection{Case Study 1: Benchmark Evaluation Results by Difficulty}

\begin{figure*}[!h]
\centering
\includegraphics[width=\textwidth]
{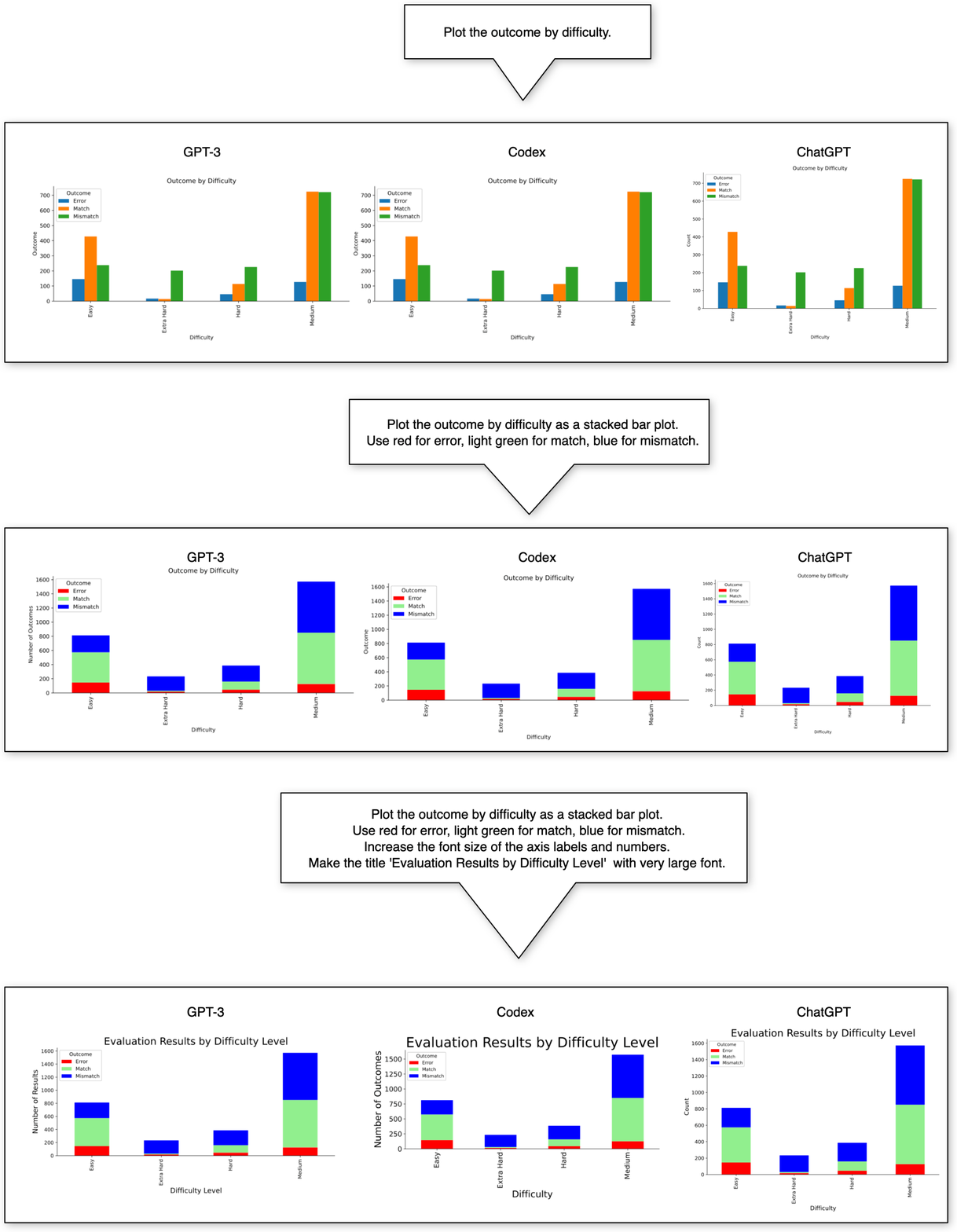}
\caption{Case Study 1: Visualisation Refinements for Benchmark Evaluation Results by Difficulty}
\label{refinements_1}
\end{figure*}

Fig. \ref{refinements_1} demonstrates building on the initial query \textit{"Plot the outcome."} using the dataset example from Fig. \ref{chat2vis_eg}. We first ask for the results summarised by difficulty.  Following the initial rendering, we suggest a stacked bar chart as an alternative plot type. We specify desired colours for each category.  Finally, we enlarge the font size for readability, and request a new very large title.  All LLMs interpret these vague font size requests uniquely, with no specific detail provided as to the exact size required.

\subsection{Case Study 2: Benchmark Evaluation Results by Outcome}
\begin{figure*}[!h]
\centering
\includegraphics[width=\textwidth]
{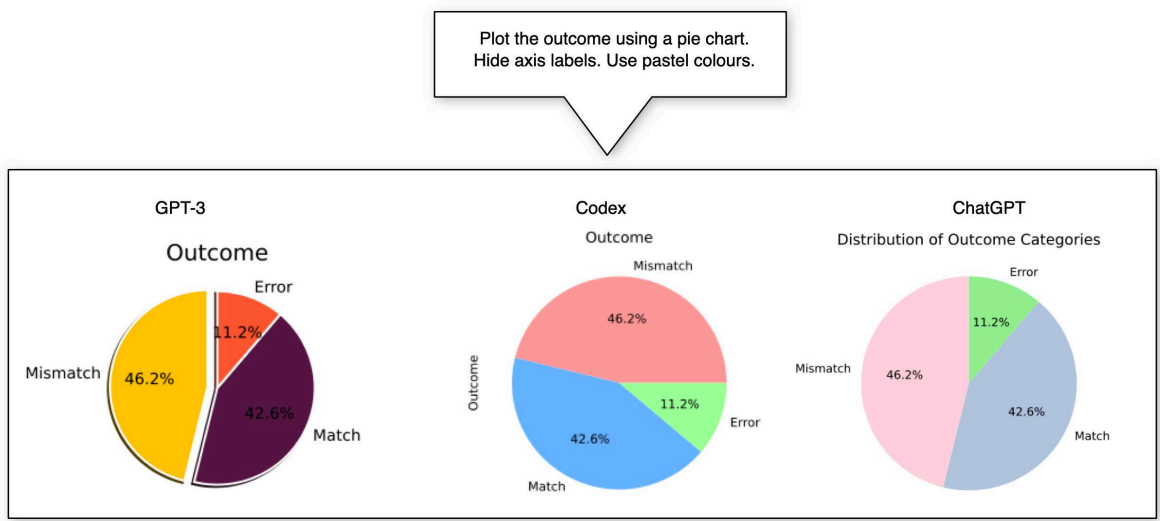}
\caption{Case Study 2: Visualisation Refinements for Benchmark Evaluation Results by Outcome}
\label{refinements_2}
\end{figure*}

Once again we begin with the base query \textit{"Plot the outcome."} from the dataset example in Fig. \ref{chat2vis_eg}. Illustrations in Fig.  \ref{refinements_2} demonstrate refining the query, by altering the chart type to a pie plot.  We make a suggestion to use pastel colours, an imprecise request interpreted differently by all LLMs. GPT-3 is considered the least accurate in its colour choice. 

In an attempt to remove the axis label \textit{"outcome"}, we use positive instruction with the keyword 
\textit{"hide"} rather than \textit{"remove"} or \textit{"delete"}. The request is adhered to by both GPT-3 and ChatGPT, but somewhat ignored by Codex.

\subsection{Case Study 3: Benchmark Evaluation Results by Database}
\begin{figure*}[!h]
\centering
\includegraphics[width=\textwidth]
{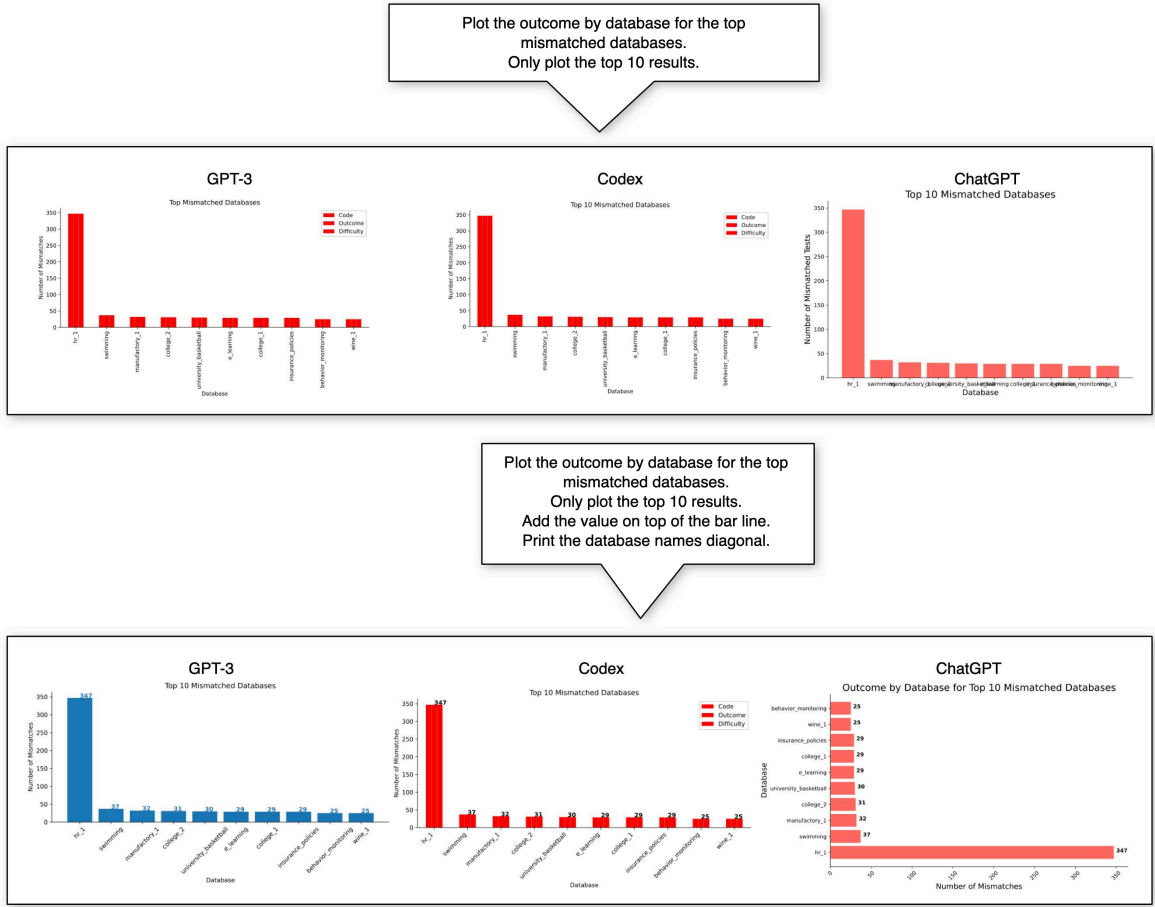}
\caption{Case Study 3: Visualisation Refinements for Benchmark Evaluation Results by Database.}
\label{refinements_3}
\end{figure*}

We highlight the ability of LLMs to categorise data based on input query. Fig. \ref{refinements_1} shows fine-tuning of the initial prompt \textit{"Plot the outcome."} from the example dataset in Fig. \ref{chat2vis_eg}. We first ask for the results categorised by database. With the large number of databases tested, we request the top 10 with mismatched outcomes.  All models visualise correct results, with GPT-3 and Codex yielding an unnecessary legend, and ChatGPT rendering unreadable database names.  Therefore, to enhance readability, we suggest diagonal alignment of the database names and integrating y-axis values above the bars.  The non-deterministic nature of the LLMs occasionally triggers variability in chart generation that we witness here. This additional instruction results in GPT-3 changing plot colour and removing its legend rendered previously.  ChatGPT alters its approach and decides on a horizontal bar plot to visualise the request.

\subsection{Case Study 4: Multilingual Requests}
\begin{figure*}[!h]
\centering
\includegraphics[width=\textwidth]
{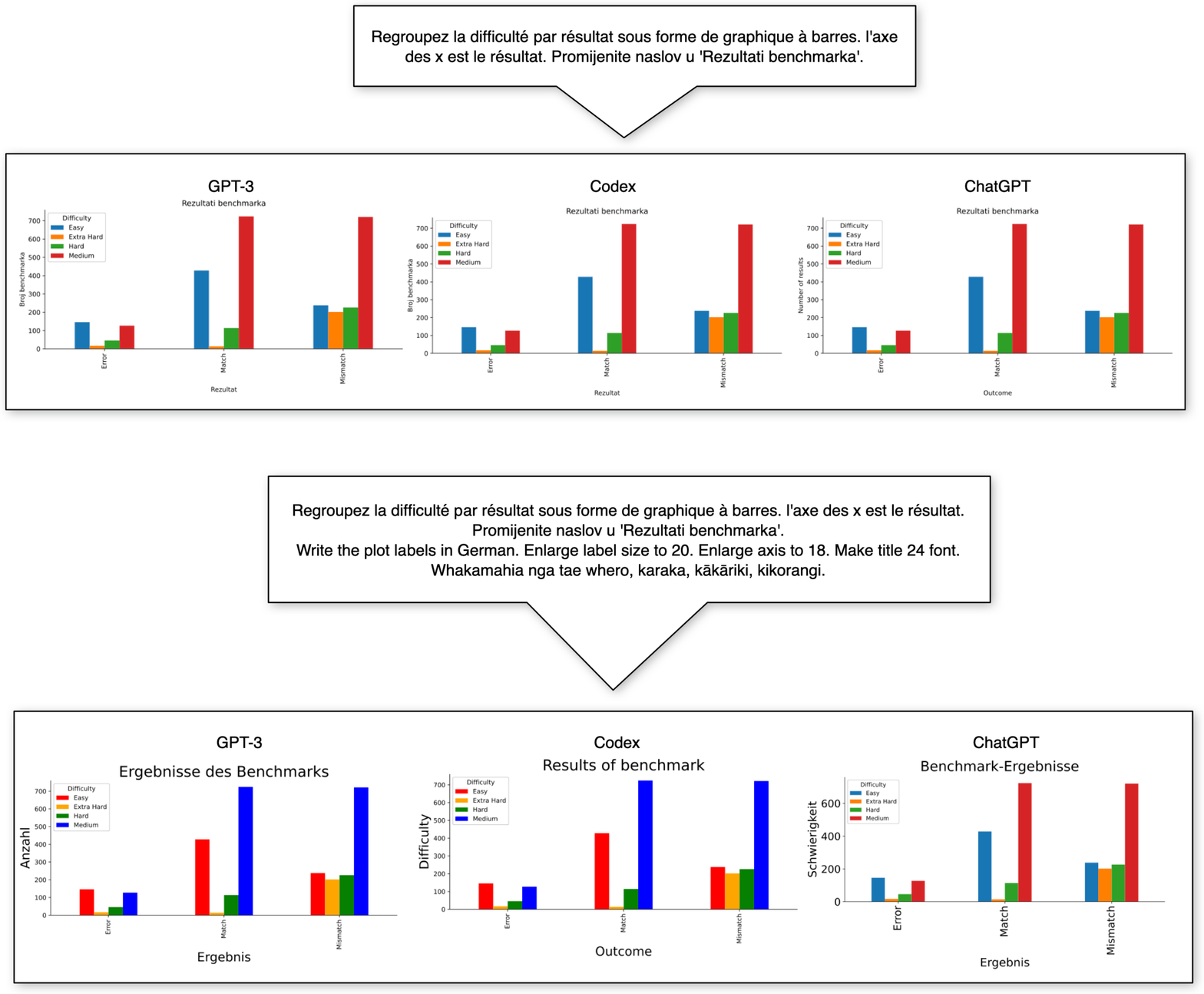}
\caption{Case Study 4: Visualisation Refinements using Multilingual Requests.}
\label{multilingual1}
\end{figure*}

In this case study we showcase the proficient capability of the LLMs to comprehend mixed multilingual requests as demonstrated in Fig. \ref{multilingual1}. We consider results categorised by difficulty, building on the initial prompt \textit{"Plot the outcome."} from Fig. \ref{chat2vis_eg}. By soliciting the outcome be plotted along the x-axis,  we submit the query using the French language \textit{"Regroupez la difficulté par résultat sous forme de graphique à barres. l'axe des x est le résultat."}, translated\footnote{via Google translate https://translate.google.com} as \textit{"Group difficulty by outcome as a bar chart. The x axis is the result."}. Additionally we ask for a title change to "Benchmark Results" using the Croatian phrase \textit{"Promijenite naslov u 'Rezultati benchmarka'."}.  Each LLM correctly renders the specified plot, with title presented in Croatian and retaining both legend and bar labels in English.   GPT-3 and Codex translate both axes' labels into Croatian, while ChatGPT preserves the English labelling. 

We refine the instructions further, using the English language to request  plot labels be expressed in German, label size increased to 20,  axis labels enlarged to 18, and an increase in title font size to 24. Asking for a change in colour ordering of plot bars, we use the New Zealand language Te Reo Māori. The appended request suggests the LLMs use colours red, orange, green and blue, submitting the phrase \textit{"Whakamahia nga tae whero, karaka, kākāriki, kikorangi."}. All 3 LLMs change font sizes of the requested elements, with ChatGPT also increasing tick label sizes. Both GPT-3 and ChatGPT provide German labels for the title and axes as requested, however Codex reverts to English labelling. GPT-3 and Codex correctly interpret the colouring request, with ChatGPT retaining its original colour scheme.

\subsection{Evaluation against nvBench}
\label{section:nvBench_eval}
Our findings in evaluating data visualisations from Chat2VIS with those from nvBench indicate a 43\% rate of success using the Codex LLM.  From the 3,003 queries tested, Fig. \ref{chat2vis_eg} shows 1,280 charts exactly matched, 1,387 mismatched, and 336 were unable to be rendered attributable to the erroneous code generated by the LLM. Fig. \ref{refinements_1} presents summary statistics of our results by difficulty level.  
To understand the limited success rate we further inspect a sample of failed instances. On examination we show discrepancies are broadly categorised into four categories, outlined in the following sections. 

\subsubsection{Benchmark Inaccuracies}
We observed instances where the Chat2VIS chart was precise in its rendering, however imperfections in the nvBench specification and inconsistency with database information resulted in mismatching output.  We illustrate in Fig. \ref{benchmark_shortcomings}(a) a discrepancy in categorical values with variant spellings of \textit{"Andreo"} and \textit{"Lo"} in the nvBench specification contrasted with \textit{"Andreou"} and \textit{"Lou"} persisted in the database.

Interestingly, we noted the presence of at least five nvBench specifications omitting the first query. On passing an empty string to Chat2VIS, the LLM actions its skilled decision making to render a visualisation of interest, as demonstrated in  \ref{benchmark_shortcomings}(b). 

In a subset of examples we found inconsistencies between the query request and the SQL denoted in the nvBench specification. Our findings unearthed instances where the SQL included an \texttt{ORDER BY} clause but the NL query omitted instructions to \textit{"order"} or \textit{"sort"} results.  Consequently the Chat2VIS visualisation accurately depicted the data points but mismatched that of nvBench's sorted chart.
    
Our methodology utilises only the query component of the specification,  disregarding other instructions contained within. Consequently an assortment of examples stored requests for grouping results inside the "binning" mark of the specification, therefore omitting to incorporate the request as part of the NL query. We illustrate in  Fig. \ref{benchmark_shortcomings}(c) a common mistake where the Chat2VIS chart summarised by date, and the equivalent nvBench specification grouped by weekday, an instruction solely provided within the "binning" mark.

Further inspection of mismatches showed periodically, without additional stipulations inside the specification, nvBench incorrectly renders only a partial result set, while Chat2VIS visualises the complete set of data. An example is depicted in Fig. \ref{benchmark_shortcomings}(d). The reasoning for such behaviour within nvBench is unknown.
Additionally, we noted for some instances executing the SQL query from the nvBench specification directly against the database yields conflicting figures to those visualised by nvBench, as shown in Fig. \ref{benchmark_shortcomings}(e). 
Furthermore we encounter truncating of numeric float types to integer values without additional stipulations inside the specification as shown in Fig \ref{benchmark_shortcomings}(f). Consequently causing additional mismatches between nvBench and Chat2VIS results.

\subsubsection {Methodology Limitations}
With the absence of a prescribed methodology within the nvBench specification to implement benchmark comparisions, utilising our own approach came with limitations.

Periodically the Chat2VIS chart was accurate visually, but our strategy for comparison did not yield expected results.  
When grouping results by category, with some categories  devoid of data, our findings showed Chat2VIS tends to return values only for the populated categories. On the contrary, the nvBench specification lists all categories, and those without data are assigned a zero value.  Despite being visually similar, the two charts differ in their underlying data points, leading to a mismatch. Fig. \ref{benchmark_shortcomings}(g) illustrates the dilemma in grouping data by week day when some days have no observations.

The nvBench queries often specify the ordering or sorting of results based on a nominated axis, either in ascending or descending order. As we do not rearrange the x and y vectors if sorting is requested, an exact match is necessary for the charts to be equivalent.  However, this approach presents a challenge when multiple x values share the same y value, rendering a correctly ordered, but not identical, chart. We illustrate this assorted ordering in Fig. \ref{benchmark_shortcomings}(h) with the majority of bars  value 1.
    
The Chat2VIS prompt assists the LLM in conforming to our expectations with Python syntax. Nonetheless, with varying charting techniques in Python, the LLM may sometimes opt for an unconventional approach to plotting. Consequently our extraction method for x and y vectors may not work as anticipated, leading to null values, and a mismatch.
    
We observed cases where missing values in numeric fields within the nvBench database tables are represented as empty strings. In these circumstances we expect NULL values. The empty strings cause Python to import the entire numeric column as a non-numeric data type. As a result, queries using numeric values to filter the non-numeric columns fail to execute.  To prevent such occurrences, enhancing the Chat2VIS interface to provide adjustments to dataset attributes may alleviate the problem.  Futhermore,
if preemptively addressed, we may have potentially mitigated these issues prior to conducting our evaluation.

On occasion we observed columns within the database holding numeric data but stored defined with a character type.  As a consequence, queries with mathematical computations, such as determining the minimum numerical value, do not always yield the expected result, minimising the string value instead.  Python is then unable to render the desired plot as it expects numeric data.
To encourage correct syntax, the Python version is denoted within the Chat2VIS prompt. Nevertheless, sporadically the LLM generates outdated syntax, leading to errors during execution.    
We also noted periodically the LLM generates a verbose script that surpasses our preset token restriction, hence execution of the truncated script results in error.

\subsubsection{Ambiguity}
Ambiguity within the query text resulted in different interpretations by the LLM and the benchmark. This culminated in the rendering of divergent charts. Hence it was challenging to determine the accuracy of either. Queries such as \textit{"...could you sort bars in desc order?"}, \textit{"...order by the bars from high to low."}, \textit{"...could you list bars in descending order?"}, and \textit{"...order by the bar in descending."}  prompted Chat2VIS to sort by y-axis values. However nvBench interpreted these instructions as sorting by the x-axis labels in alphabetical order. We illustrate an example in Fig. \ref{benchmark_shortcomings}(i).

The primary purpose of a bar chart is to facilitate the comparison of values among various groups. Typically it is uncommon to use this type of chart to represent the distribution of two text columns, however we observed such instances.
Fig. \ref{benchmark_shortcomings}(j) shows how nvBench constructs a requested chart for such an example. Meanwhile Chat2VIS infers a more appropriate visualisation (although aesthetically unappealing) of the requested information providing counts for each category.

\subsubsection{Query Misinterpretation} 
Occasional misinterpretation by the LLM was evident, prompting Chat2VIS to generate an erroneous visualisation.  Date and time values can be problematic. Within Python they are represented together by the datetime object. To group information by date, we disregard the time component. 
We observed nvBench correctly utilising the date component only, hence accurately grouping datetimes with the same date but different timestamps into one category.
However, Chat2VIS often considered both the date and time component and as a result datetimes were incorrectly grouped individually.

The methodology used to build nvBench periodically elicits large groups of analogous benchmark examples. Our findings showed filter request such as \textit{"...commission is not null or department number does not equal to 40..."}, was present in over 70 examples. Unfortunately it was misunderstood by Chat2VIS. The inadvertent use of the \textit{"and"} operator instead of \textit{"or"} in all instances augmented overall mismatches.

We encountered one request in the Chinese language represented in unicode characters. Unfortunately Chat2VIS misunderstood the request as a result of  incorrectly importing the characters from the nvBench specification.  Furthermore, 
we noted in some instances, without direction, the LLM incorrectly self-imposed a restriction on the quantity of returned results. In future, including instruction within the prompt of the LLM would avoid such truncation.
However, in spite of the LLMs' adept aptitude for skilfully generating Python code from natural language, occasionally, without explanation, its interpretation was erroneous.

\subsection{Evaluation against nlvUtterance}
\label{section:nlv_corpus_eval}
\begin{figure}[!htb]
\centering
\includegraphics[width=3in]
{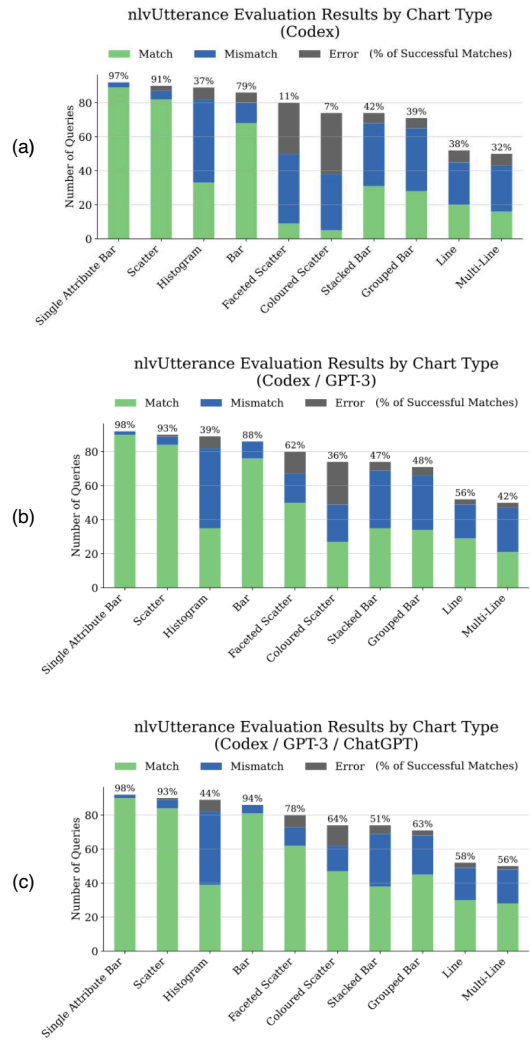}
\caption{nlvUtterance Evaluation Results.}
\label{nlv_corpus}
\end{figure}

Our evaluation of Chat2VIS using the nlvUtterance benchmark dataset shows a 50\% rate of success over all chart types solely using Codex, 63\% success of at least one of Codex and GPT-3 generating a matching chart, and 72\% success of at least one of Codex, GPT-3, and ChatGPT generating a matching chart.  Fig. \ref{nlv_corpus} depicts evaluation results categorised by chart type for each stage of testing.  Single-attribute bar plots and scatter plots were the most-represented chart types in the dataset, and both were found to have the highest success rates. Evaluations using Codex shows in Fig. \ref{nlv_corpus}(a) coloured scatter plots and faceted scatter plots were found to have significantly lower success rates, however outcomes were more favourable once generated via GPT-3 Fig. \ref{nlv_corpus}(b) and ChatGPT Fig. \ref{nlv_corpus}(c). 

It is important to note that differences observed in the rendered charts between Chat2VIS and nlvUtterance do not necessarily imply inaccuracy. In some cases the generated plot correctly displayed the requested information, but in an alternative charting configuration attributable to the lack of specification and ambiguity within the query. Nevertheless, without direction of a defined measurement methodology within the benchmark, our evaluation often deemed such discrepancies a mismatch.  On further examination of mismatched samples, we explain the higher rates of failure for grouped and stacked bar charts, histograms, colour and faceted scatter charts, and single and multi-line charts.  Fig. \ref{nlvUtterance_egs} illustrates a sample of visualisations generated for the ten chart types based on the movies dataset, enabling comparisons with those presented in prior work \cite{Srinivasan2021Utt}.

\subsubsection{Bar Charts}
 A substantial number of grouped bar charts were classified as a mismatch notwithstanding correct presentation of the requested data. Considering the query \textit{"average production budget by creative type and content rating"}, the benchmark arranged results grouped by content rating using colour coding to represent the creative type. However, periodically the Chat2VIS counterpart inversely grouped results by creative type with colours representing content rating.  Similar issues were observed with mismatches  between stacked bar charts. Furthermore, our findings showed instances where benchmark stacked bar charts were presented by Chat2VIS as grouped bar charts, accurately conveying the information, but not in accordance to the benchmark standard.  Should we have deemed the visualisations as correct, as indeed they were despite deviating from the benchmark, the rate of success would increase significantly.  Nonetheless,  the absence of methodology within the benchmark failed to provide guidance in such circumstances.

\subsubsection{Histograms}
We refrained from providing explicit instructions to the LLM on which chart type to render. Consequently a significant number of benchmark histograms were instead plotted as bar charts. 
    Queries such as \textit{"How many orders were placed for each order quantity?"} and \textit{"show me a bar chart of count by order quantity"} did not imply the data should be represented as a histogram, and hence the LLM decided the most appropriate representation of the data was in the form of a bar chart.

Furthermore, queries neglected to provide information pertaining to binning size, and consequently the LLM's decision often conflicted with benchmark visualisations.

\subsubsection{Scatter Charts}
Colour scatter charts represent categories using contrasting colours within a single plot.  Codex tended to neglect colour coding categories, and produced single-coloured charts.  
GPT-3 did not show such limitations, however, a large percentage of both Codex and GPT-3 mismatches were attributed to the generation of incorrect programming syntax. The \textit{"c"} parameter value of the Python plotting function expects to receive colour information for chart aesthetics and to distinguish categories. However the LLMs repeatedly set it to enumerated variable values, resulting in the execution of erroneous scripts.  ChatGPT was inclined to correctly assign this function parameter leading to a more successful outcome. 

Faceted scatter charts extract categories presented in a coloured scatter plot and render them as separate charts. However on further inspection of the queries, we noted instances lacking in instruction to suggest a faceted chart. Consequently we decided to also accept single scatter plots categorised by colour.
Nevertheless, as with coloured scatter charts, Codex often neglected to distinguish the categories using colour coding, and proceeded to incorrectly represent all points in a single colour.  Neither GPT-3 nor ChatGPT demonstrated such limitations, achieving a higher percentage of successful outcomes. However, as previously demonstrated, owing to the absence of benchmark methodology, we resolved to accept these alternative charts. 
If we had not done so the rate of success may have been affected.

\subsubsection{Line Charts}
The least-represented chart types in the dataset are single and multi-line plots.  The most common cause of mismatch was the LLM selecting to render the information as a bar chart. However, although it still accurately presented the requested information when a line chart was not explicitly requested, 
it was not in accordance with benchmark specifications. In addition, Chat2VIS multi-line plots on occasion inversely rendered the x-axis and line colour categories compared to that of nlvUtterance, hence unsuccessful in meeting benchmark standards. Once more, these decisions of determining if benchmark standards are met significantly impact culminating a successful outcome.

\subsection{Comparison With Previous Work}
\label{section:comp_prev_work}
In contrasting our nvBench benchmark findings with those presented in previous work it should be noted that our sample set differs from test sets used to evaluate approaches developed within other studies.  
Overall accuracy of existing NL2VIS systems  employing  nvBench as a benchmark have previously \cite{song2022rgvisnet} been contrasted with Seq2Vis, Transformer, ncNet and RGVisNet  as noted in Table \ref{table:nvBench_Results}.

\begin{table}[h!]
\centering
\caption{nvBench Performance Comparison.}   
\begin{tabular}  {lc}
\hline 
System & Accuracy \\
\hline
Seq2Vis & 2\% \\
Transformer & 3\%\\
ncNet & 26\%\\
RGVisNet & 45\%\\
\textbf{Chat2VIS}&\textbf{43}\% \\
 \hline\end{tabular}
\label{table:nvBench_Results}
\end{table}

Table \ref{table:nlvcorpus_Results} summarises our findings using the nlvUtterance benchmark, with those from NL4DV evaluated on the benchmark in previous studies \cite{Srinivasan2021Utt}. It should be noted the NL4DV results are based on a 755 instance dataset pertaining to singleton query sets only. In our study we included all query sets, with the exclusion of those pertaining to the two omitted line charts.

\begin{table}[h!]
\centering
\caption{nlvUtterance Performance Comparison.}   
\begin{tabular} {lc}
\hline 
System & Accuracy \\
\hline
Chat2VIS (Codex) & 50\% \\
Chat2VIS (Codex/GPT-3) & 63\%\\
Chat2VIS (Codex/GPT-3/ChatGPT) & 72\%\\
NL4DV & 64\%\\
 \hline\end{tabular}
\label{table:nlvcorpus_Results}
\end{table}

\section{Discussion}

The case studies presented confirm the skills and adaptability of the LLMs to interpret multilingual requests, including mixed language queries. Our study verifies the capabilities of the LLMs to label plot elements in accordance with language preference. 

Furthermore, we show with limited instruction in the initial query, the LLM decides on plot aesthetics.  In addition, our case studies attest to the unique ability of Chat2VIS to refine chart elements further by appending additional instruction to the initial query.

Our benchmarking evaluations confirm how two baselines employed to evaluate Chat2VIS in this study are as yet unable to fulfil all the quality characteristics \cite{Kounev2020Benchmark} 
of \textit{reproducibility}, \textit{fairness}, and \textit{verifiability}.  There is an abundance of diverse visualisation elements, aesthetics, and chart styles, from an variety of available programming language libraries.  This raises difficulties in generating \textit{reproducibile} and measurable outcomes from natural language queries.  The predominant use of vega-lite specifications in current benchmarking studies \cite{Srinivasan2021Utt}\cite{luo2021nvbench} periodically separates important visualisation information from the natural language query. This limits the effectiveness of other NL2VIS architectures, questioning \textit{fairness}.  Inconsistencies between test case data and visualisation outcomes raise concerns to the \textit{verifiability} of the benchmark examples.  

We detected oversights in the nvBench data specifications and corresponding visualisations.  We note the assertion that the benchmark is validated by 23 experts and 300+ crowd workers\cite{nvBench_SIGMOD21}.  We attribute these inconsistencies to 
 the inherent human fallibility of manual evaluations, which contrasts with the higher accuracy of automated processes.

From this study we showed current benchmarks are vague in their descriptions, lending to subjective interpretations of measurement methodology.  Subsequently, without the guidance to implement standards and evaluate results, we generate inconsistent quantitative evaluation metrics with previous studies.

In future, to establish a robust benchmarking tool for NL2VIS, we foresee a collection of visualisations for each natural language query.  Coupled with this would be a comprehensive methodology for chart comparison and evaluation.  The strategy must be autonomous of chosen charting frameworks.  Nevertheless, current benchmarking processes have followed an inverse approach.  Initially building a set of pre-designed charts, they subsequently yield a collection of queries which potentially generate the desired visualisations.

\section{Conclusion}
 This study presented the novel features of Chat2VIS for converting natural language text into data visualisations using large language models.  It has contributed to the gap in existing literature where no other NL2VIS solution can refine chart aesthetics to such an  extent.  Moreover, its proven ability to comprehend multilingual requests has addressed a shortfall in prior approaches. 

 We presented our innovative automated approach to quantitative benchmarking of NL2VIS. Exploring the challenges and requirements to robustly compare results has tackled the dearth in literature within this field. We have identified areas of concern and offered improvements in baseline standards so we can foster advancements within NL2VIS.


%


\newpage
\appendix

\section{Chat2VIS Interface}
\begin{figure*}[!h]
\centering
\includegraphics[width=\textwidth]
{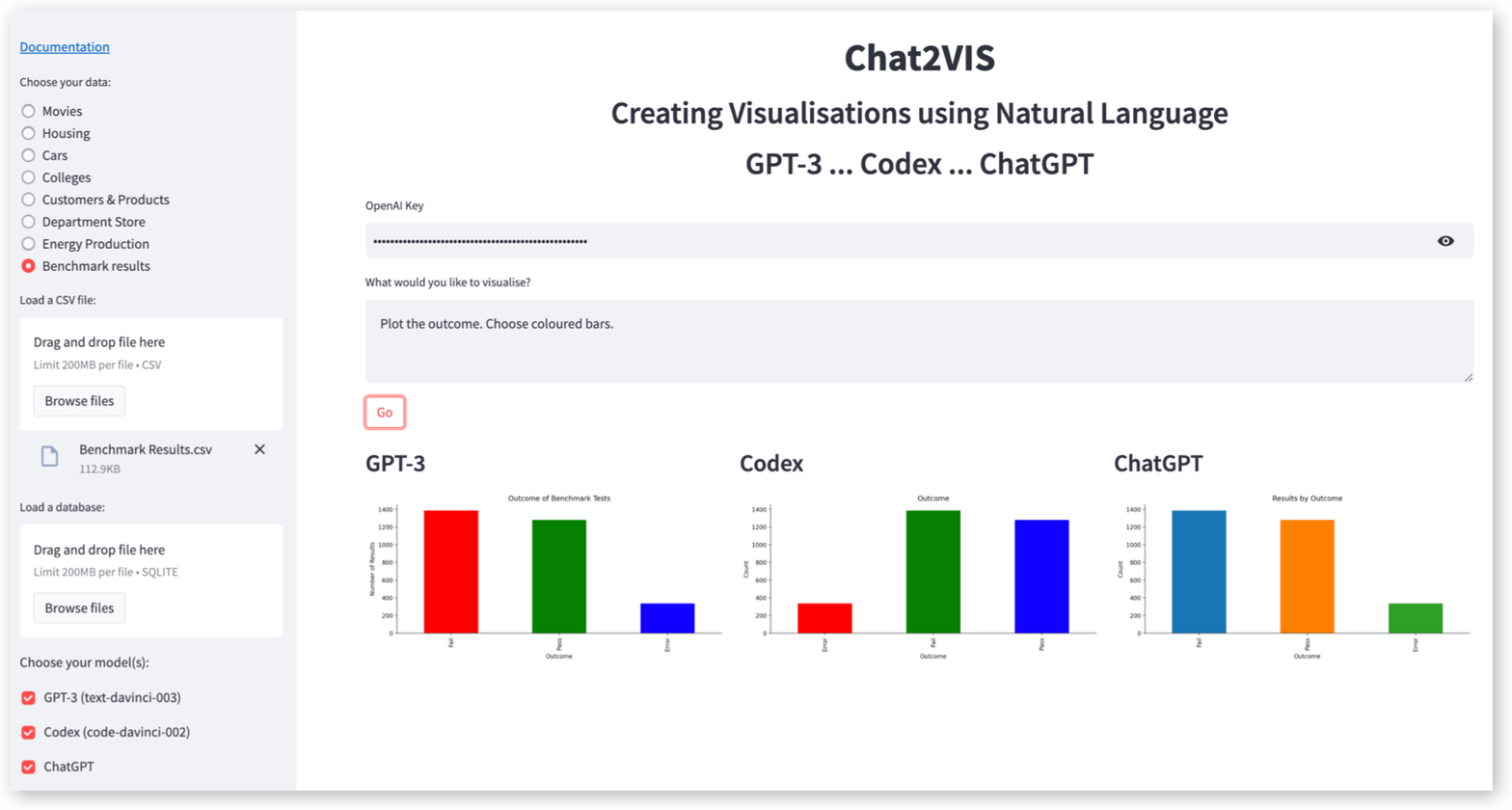}
\caption{Chat2VIS Interface}
\label{interface}
\end{figure*}

\section{Prompt Engineering}
\begin{figure*}[!h]
\centering
\includegraphics[width=\textwidth]
{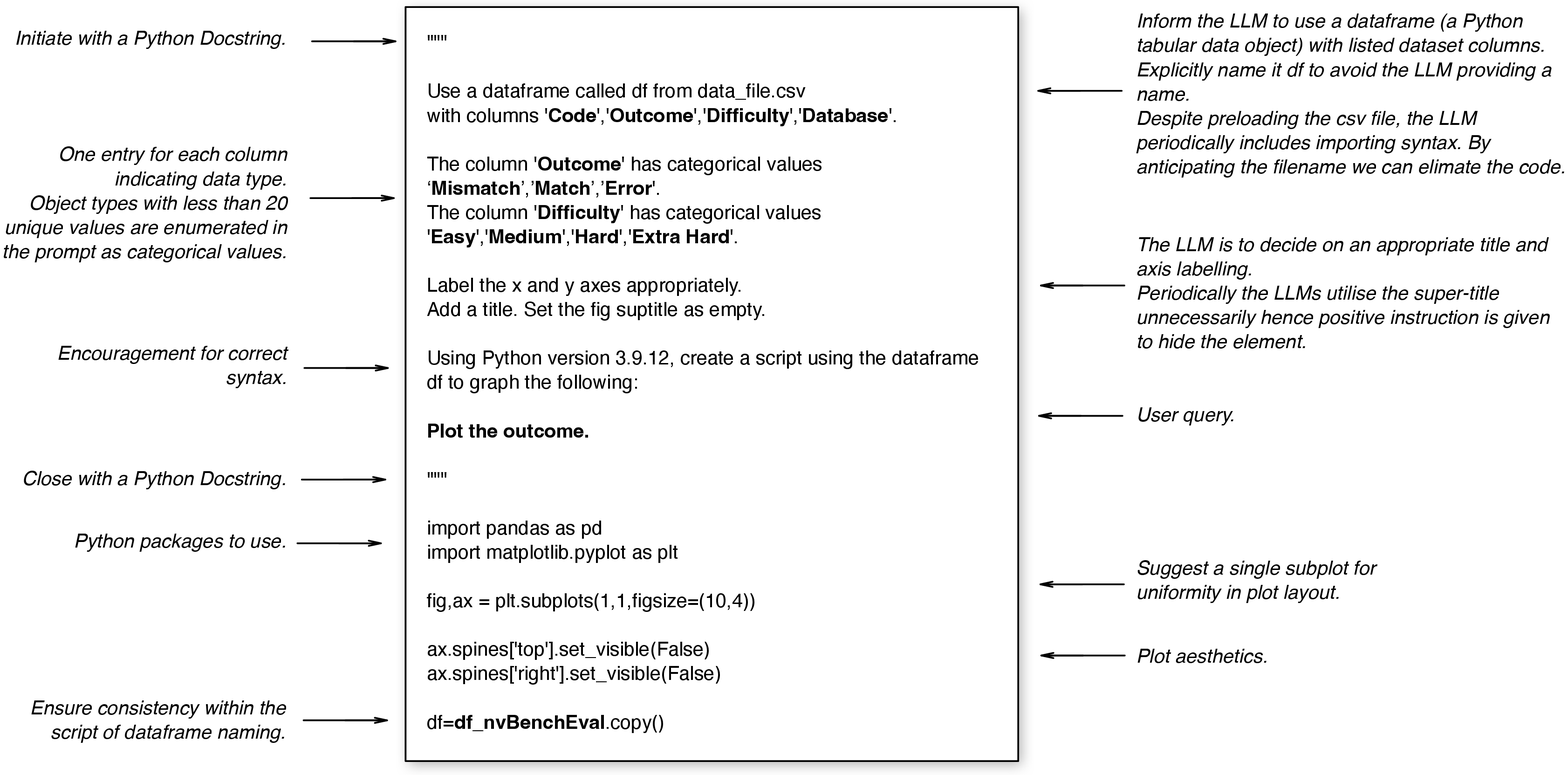}
\caption{Explanation of Chat2VIS Prompt}
\label{chat2vis_prompt}
\end{figure*}

\section{nvBench Testing Methodology}
\label{nvBench_method}
The testing methodology for nvBench as illustrated in Fig. \ref{testing_overview} is described as follows:
\begin{figure*}[!h]
\centering
\includegraphics[width=\textwidth]
{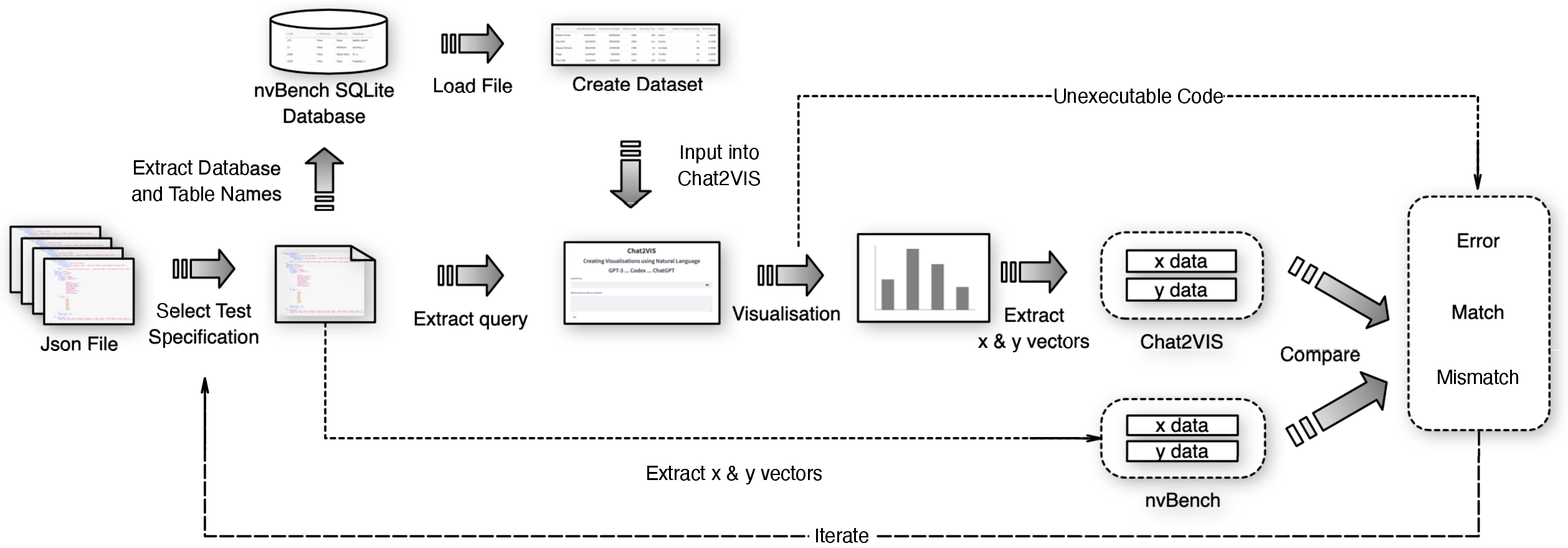}
\caption{Overview of nvBench Benchmark Testing.}
\label{testing_overview}
\end{figure*}

\begin{enumerate}
\item Select a json test specification from the nvBench test set. 
\item Extract the database name from the db\_id mark, and the table name specified in the VQL mark following the FROM keyword.
\item Import the SQLite database table into a Python Pandas dataframe structure.  
\item Extract the first query from the  nl\_queries mark.
\item Submit the query and dataset to Chat2VIS opting for the Codex model, and render the requested visualisation.
\item Document the outcome as an Error should the code fail to execute.
 \item Construct vectors of the x and y data coordinates by extracting the Chat2VIS chart elements.
 \item Construct vectors of the x and y data coordinates within the x\_data and y\_data marks of the  nvBench specification.
\item Apply adjustments to the vectors to address complications impacting a successful comparison: 
    \begin{itemize}
        \item Ensure naming consistency of calendar units, such as recasting Tue, Tues, and tuesday as Tuesday; Sept, Sep september as September.
        \item Sort by ascending y values if keywords "sort" and "order" are not specified.
        \item Cast integer values to floats, and round all floats to 5dp.
    \end{itemize}
 \item Compare Chat2VIS and nvBench vectors. A precise match marks the outcome as a "Match", else it is marked as a "Mismatch". 
\end{enumerate}

\section{Benchmark Examples}

\begin{figure*}[!h]
\centering
\includegraphics[width=\textwidth]
{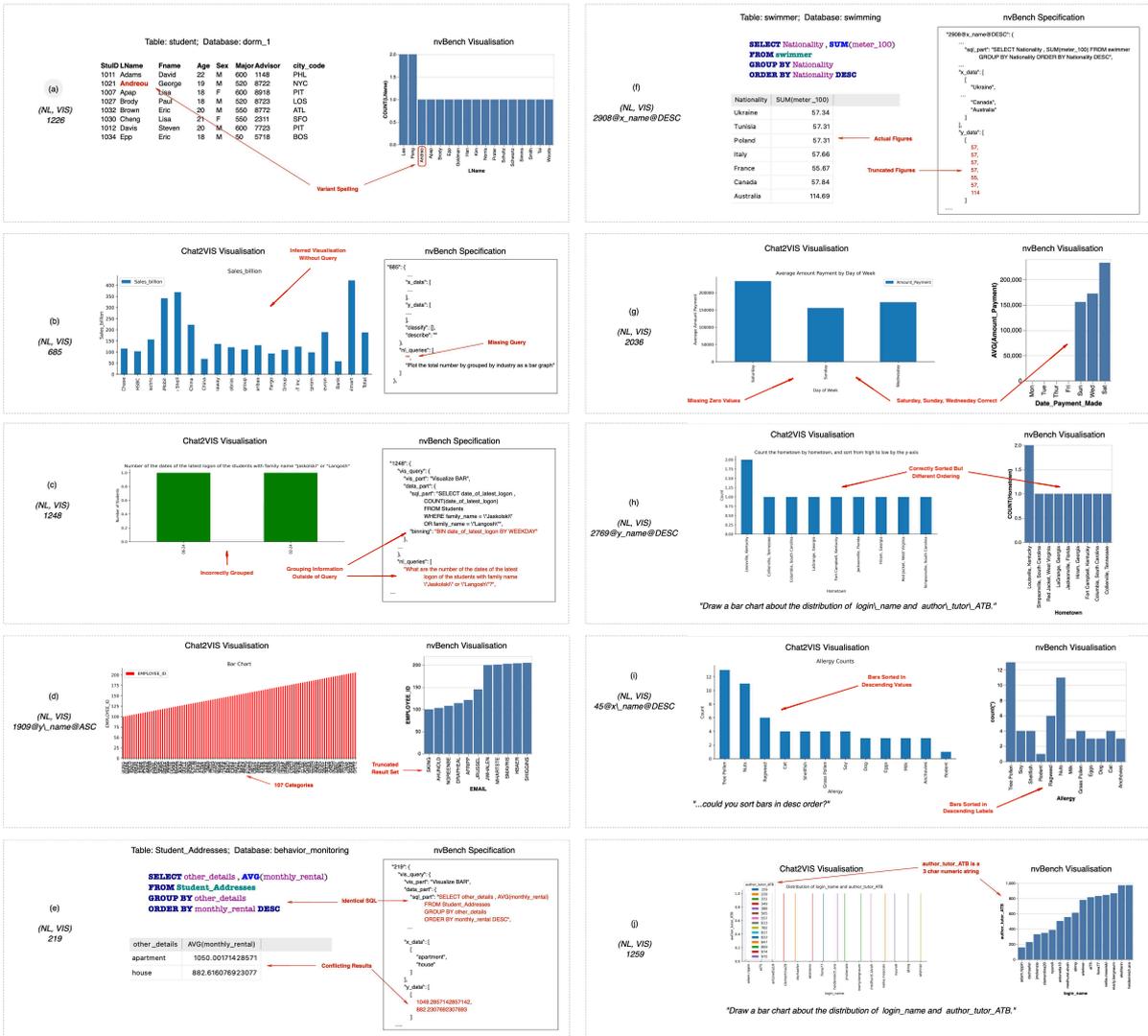}
\caption{Examples of Mismatched Visualisations Between Chat2VIS and nvBench}
\label{benchmark_shortcomings}
\end{figure*}


\begin{figure*}[!h]
\centering
\includegraphics[width=\textwidth]
{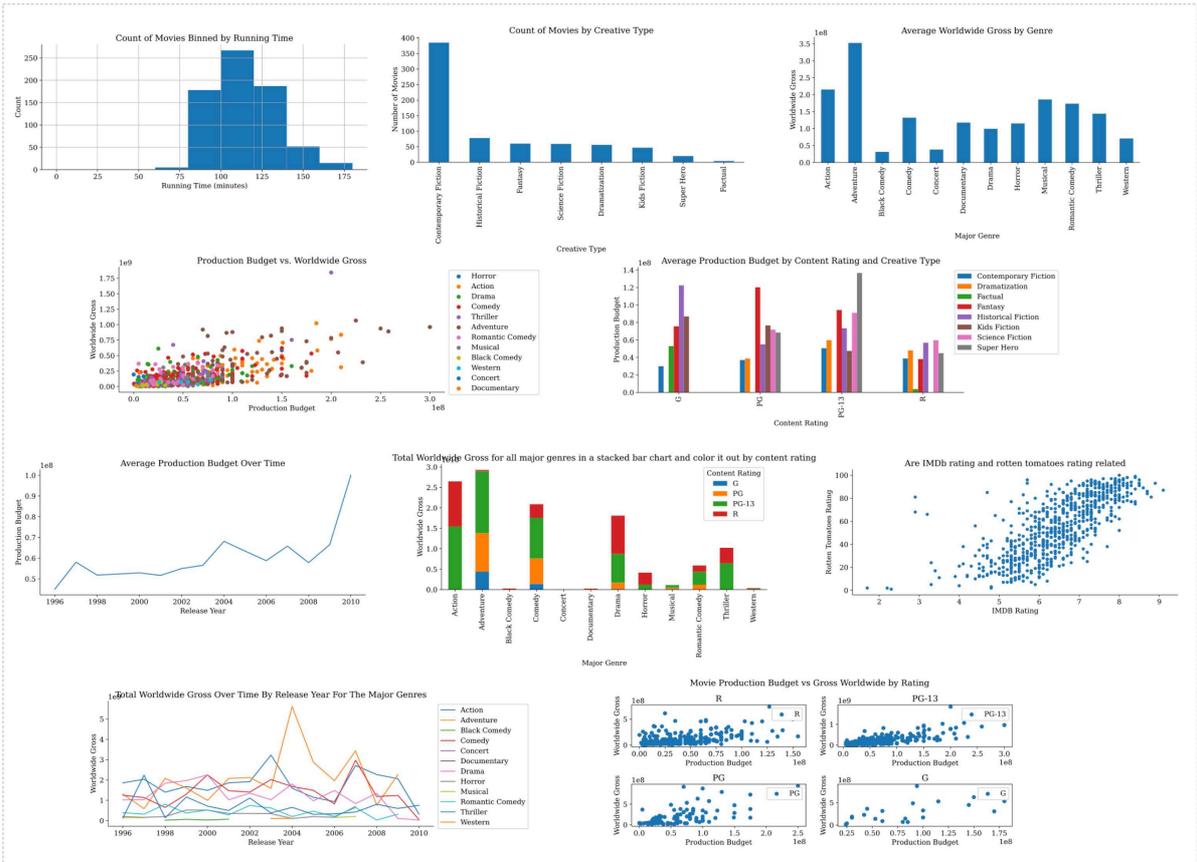}
\caption{nlvUtterance Movies Dataset Examples from Chat2VIS}
\label{nlvUtterance_egs}
\end{figure*}

\end{document}